# Extremal Properties of Random Systems

L. Frachebourg, I. Ispolatov, and P. L. Krapivsky

*Center for Polymer Studies and Department of Physics*

*Boston University, Boston, MA 02215, USA*

(July 5, 1995)

We find that the probability distribution for the largest intervals $p(l)$ exhibits universal properties for different systems including random walk and random cutting models. In particular, $p(l)$ has an infinite set of singularities at $l = 1/k$ with $k = 2, 3, \ldots$ which become weaker and weaker as $k \to \infty$; additionally, $p(l)$ has an essential singularity at $l = 0$. These properties are found also in many dimensional situation.

PACS numbers: 05.40.+j

Systems with complex landscapes are of great current interest in a variety of fields ranging from optimization to evolution [1]. The phase space in such systems is usually broken into many valleys. The multivalley structure determines many peculiar properties as it was first realized for spin glasses [2]. The large valleys fluctuate from sample to sample and lead to anomalous long-time behavior, ergodicity breaking, etc. Large valley distribution and other extremal characteristics play therefore a significant role in various problems.

To study extremal properties of random systems, we consider two very simple geometrical models. The first model arose as a byproduct of our previous work [3] which studies the kinetics of annihilation of charged particles in $1D$ two-component plasma with a truly $1D$ Coulomb interaction and neutrality condition. This system is decoupled into a succession of uncharged chains which evolve independently [3]. The long time behavior is therefore determined by long chains, in particular the time of the complete annihilation is directly related to the *longest* uncharged chain. Thus we run into a problem of the longest uncharged chain distribution which exhibits a surprisingly rich behavior. Although some of the features are model-dependent, the basic properties of the longest interval distribution function such as the existence of an infinite countable set of singularities and their locations seem universal. At least, we will demonstrate that similar features arise in another very different model, namely the random cutting model. We therefore anticipate that our qualitative conclusions should provide an insight on the probability distribution of the largest valley in disordered systems and systems with a complex landscape.

In the first problem, we consider sequences of $2N$ charges $Q_j$, $Q_j = \pm 1$, with $\sum Q_j \equiv 0$ to satisfy the neutrality condition. It is clear that such a sequence can be visualized as a random walk (RW) starting at the origin at time $t = 0$, jumping a step up at $t = j$ if $Q_j = 1$ and a step down if $Q_j = -1$, and returning back to the origin at $t = 2N$ (Fig. 1). A RW can return to the origin few more times, and we ask for the probability distribution of the *longest* time interval, $2L$, between successive departures and arrivals (see Fig. 1).

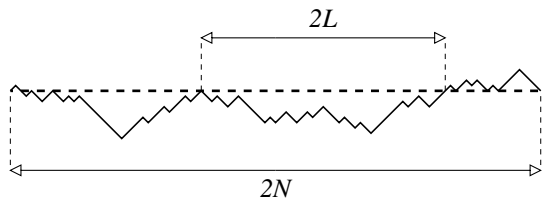

FIG. 1. A typical random walk with $2N = 100$ and $2L = 46$

A typical number of returns scales as $\sqrt{N}$, and so does a typical time interval. However, the longest time interval scales as $N$ rather than $\sqrt{N}$. To demonstrate this and other peculiar properties of the longest time interval let us calculate the total number, $T_N(L)$, of RWs with the longest time $2L$. If $2L \geq N$, a time interval of length $2L$ is always the longest one and therefore $T_N(L)$ is easily computed:

$$T_N(L) = \sum_{K=0}^{N-L} R_K F_L R_{N-K-L}. \qquad (1)$$

Here $R_K = (2K)!/K!K!$ gives the total number of "head" RWs of length $2K$ from the origin $(t,x) = (0,0)$ to the beginning of the longest interval $(2K, 0)$, $F_L = 2(2L-2)!/(L-1)!L!$ gives the total number of RWs starting at $(2K, 0)$ and then *first* returning to $x = 0$ at $t = 2K + 2L$, and $R_{N-K-L}$ gives the total number of "tail" RWs of length $2N - 2K - 2L$ [4]. Computing the sum in Eq. (1) yields $T_N(L) = F_L \cdot 2^{2N-2L}$ and thus the longest interval probability, $P_N(L) \equiv T_N(L)/R_N$, is equal to $F_L \cdot 2^{2N-2L}/R_N$ for $2L \geq N$. In the continuum limit $N \to \infty$ it is convenient to use the probability density $p(l) \equiv NP_N(L)$ where $l = L/N$ is the scaled length. Our previous result can then be rewritten as

$$p(l) = \frac{1}{2l^{3/2}}, \quad \text{for} \quad \frac{1}{2} \leq l \leq 1. \tag{2}$$

When $2L < N$, we should modify Eq. (1) to guarantee that the head and tail regions do not contain intervals of length $> 2L$. In the continuum limit, the probability distribution $p(l)$ obeys the integral relation:

$$p(l) = \int_0^{1-l} \frac{f(l,x)}{\sqrt{\pi x}} \frac{1}{2l^{3/2}} \frac{f(l, 1-l-x)}{\sqrt{\pi(1-l-x)}} dx. \tag{3}$$

Here $x \equiv K/N$ and $1-l-x$ are the scaled lengths of the head and tail regions, $f(l,x)$ and $f(l, 1-l-x)$ are the probabilities that the longest intervals in the head and tail regions are smaller than $l$. For example,

$$f(l,x) = 1 - \int_l^x p\left(\frac{\xi}{x}\right) \frac{d\xi}{x} \equiv 1 - \int_{l/x}^1 p(\eta) d\eta, \tag{4}$$

when $x > l$ while for $x \leq l$ we have $f(l,x) \equiv 1$. Besides $f$-subfactors, the first and the last factors inside the integral in Eq. (3) are just the continuum representation of $R_K$ and $R_{N-K-L}$, respectively.

Notice that $f(l,x)$ depends on the the tail of the probability distribution $p(\eta)$ for $\eta \geq l/x \geq l/(1-l)$. In particular, when $l \geq \frac{1}{2}$, $f$-subfactors are trivial, $f(l,x) = f(l, 1-l-x) = 1$ and Eq. (3) reduces to Eq. (2). Similarly, in the interval $\frac{1}{3} \leq l \leq \frac{1}{2}$ we readily compute $f$-subfactors from the knowledge of the probability distribution in the preceding interval, Eq. (2). After elementary computations we find

$$p(l) = \frac{1}{\pi l^{3/2}} \left(\pi - \sin^{-1}\left(\frac{3l-1}{1-l}\right) - 2\sqrt{\frac{1-2l}{l}}\right), \tag{5}$$

for $\frac{1}{3} \leq l \leq \frac{1}{2}$.

These results can in principle be generalized for any $\frac{1}{k+1} \leq l \leq \frac{1}{k}$ but even for $k=3$ the outcome is rather cumbersome. Let us just stress the most important feature: Since the expressions for the probability distribution function $p(l)$ are different in different regions $\frac{1}{k+1} \leq l \leq \frac{1}{k}$, singularities are expected at $l = 1/k$. These singularities, however, become weaker as $k$ increases. Expanding expressions (2) and (5) near $l = \frac{1}{2}$ one finds that they differ by a term $p_{\text{sing}}(l) \sim |l - \frac{1}{2}|^{3/2}$. Similarly, expanding (5) and $p(l)$ in the next region $\frac{1}{4} \leq l \leq \frac{1}{3}$ near $l = \frac{1}{3}$ one gets $p_{\text{sing}}(l) \sim |l - \frac{1}{3}|^3$. The nature of singularities then becomes clear,

$$p_{\text{sing}}(l) \sim \left|l - \frac{1}{k}\right|^{3(k-1)/2}. \tag{6}$$

Even the first singularity at $l = \frac{1}{2}$ is hardly visible (see Fig. 2). However, numerical differentiation of the data allows us to detect it.

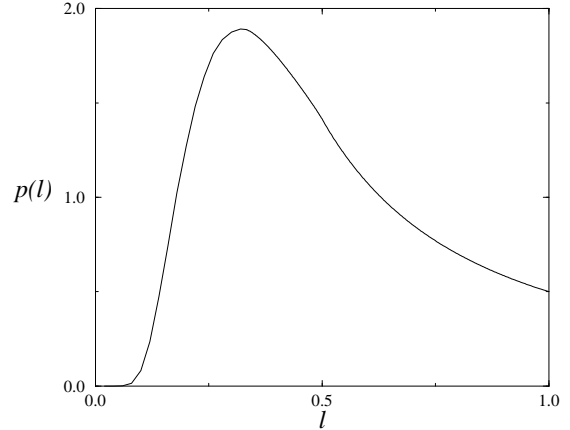

FIG. 2. Continuous probability distribution $p(l)$ for the RW problem. Numerical simulations of $10^8$ different configurations of length $2N = 1000$. Values of $p(l)$ are given by Eq.(2) for $l \geq 1/2$ and Eq.(5) for $1/3 \leq l < 1/2$.

Our numerical results also indicate that the probability density has an essential singularity $p(l) \sim \exp(-\text{const}/l)$ when $l \to 0$. To explain this behavior heuristically note that if the longest time interval between the successive returns of a RW to the origin is $2L$, this RW typically stays within a region $(-\sqrt{L}, \sqrt{L})$. The probability of that can be estimated by solving the diffusion equation in a cage with adsorbing boundary conditions. In a cage of size $B$ this survival probability behaves as $\exp(-t/B^2)$. In our case $t = N$, $B^2 \sim L$ and thus $p(l) \sim \exp(-\text{const}/l)$.

Notice that a closely related problem has been recently investigated [5]. In that model, the maximal-length chains were different from ours (just uncharged, not necessarily starting at the origin, and making arbitrary number of returns to the starting point). Numerical simulations [5] demonstrate a singularity at $l = \frac{1}{2}$, even more sharp than ours. The next singularity at $l = \frac{1}{3}$ has not been notified. However, all the singularities do exist and are located at the very same places, $l = \frac{1}{k}$, as in our model. The singularities just become weaker and weaker ($p_{\text{sing}}(l) \sim |l - \frac{1}{k}|^{k-1}$ seems possible) and hence are hardly visible for $k \geq 3$.

We now turn to the random cutting model which is defined as follows. Starting with the unit interval we cut it (the cutting probability is assumed to be uniform). Two intervals are formed, left and right, and in the next stage we cut the right one only. The cutting process is then continued with the right interval from the preceding stage. We again wish to determine the probability density, $p(l)$, for the longest interval which is formed in the cutting process.

When $l \geq \frac{1}{2}$, the probability density satisfies an integral equation

$$p(l) = 1 + \int_0^{1-l} \frac{1}{1-x} p\left(\frac{l}{1-x}\right) dx. \tag{7}$$

Indeed, we can get the desired interval of length $l$ at the first stage of the process (since $l \geq \frac{1}{2}$ we never get a piece of length $l$ or larger at any further stage). This gives the first constant contribution to the right-hand side of Eq. (7). A complimentary possibility means that at the first stage we cut at some point $x$ with $x < 1 - l$. The probability density to get the desired interval of length $l$ from remaining interval of length $1 - x$ is given by $\frac{1}{1-x} p\left(\frac{l}{1-x}\right)$, where the prefactor provides a proper normalization.

In general, the governing integral equation becomes

$$p(l) = 1 - \int_l^{1-l} p\left(\frac{x}{1-l}\right) \frac{dx}{1-l} + \int_0^l p\left(\frac{l}{1-x}\right) \frac{dx}{1-x}. \quad (8)$$

The last integral term is built as previously up to a small correction – the upper limit is now $l$ to guarantee that the first cutting gives the interval $x$ less than $l$. The two first terms again describe the situation when we get the $l$-length interval at the first stage and guarantee that in the following cutting of the remaining interval of length $1-l$, we never get an interval of length $l$ or larger. For $l \geq \frac{1}{2}$, Eq. (8) is identical to Eq. (7) if we agree to put $p(x) \equiv 0$ for $x > 1$.

Changing variables inside the integrals, $y = \frac{x}{1-l}$ in the former and $z = \frac{l}{1-x}$ in the latter, simplifies Eq. (8) to

$$p(l) = 1 - \int_{\frac{l}{1-l}}^{1} p(y) dy + \int_l^{\frac{l}{1-l}} p(z) \frac{dz}{z}. \quad (9)$$

Differentiating Eq. (9) we finally reduce it to the difference-differential equation:

$$\frac{d}{dl} p(l) = \frac{1}{l(1-l)^2} p\left(\frac{l}{1-l}\right) - \frac{p(l)}{l}. \quad (10)$$

When $l \geq \frac{1}{2}$ we have $dp/dl = -p/l$ with the boundary condition $p(1) = 1$. This is readily solved to find

$$p(l) = \frac{1}{l}, \quad \text{for} \quad l \geq \frac{1}{2}. \quad (11)$$

If $\frac{1}{3} \leq l \leq \frac{1}{2}$, the variable $l/(1-l)$ belongs to the region $[\frac{1}{2}, 1]$. Thus $p\left(\frac{l}{1-l}\right) = \frac{1-l}{l}$, and Eq. (10) becomes a simple differential equation which is solved to find

$$p(l) = \frac{1}{l} + \frac{1}{l} \ln\left(\frac{l}{1-l}\right), \quad \text{for} \quad \frac{1}{3} \leq l \leq \frac{1}{2}. \quad (12)$$

It is possible to derive recursively exact expressions for the probability density in consecutive regions $\frac{1}{k+1} \leq l \leq \frac{1}{k}$. The main conclusion remains the same: There are singularities at $l = \frac{1}{k}$ which become weaker and weaker as $k$ increases. More precisely,

$$p_{\text{sing}}(l) \sim \left| l - \frac{1}{k} \right|^{k-1}. \quad (13)$$

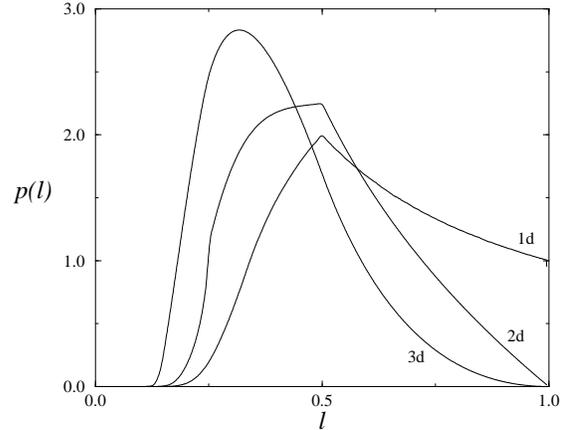

FIG. 3. Probability distribution $p(l)$ for the cutting process in $d = 1, 2, 3$. Numerical simulations of $10^9$ processes.

Consider now the small-$l$ behavior. Our simulation results indicate an essential singularity at $l = 0$. Let us try the same type of singularity, $p(l) \sim \exp(-a/l)$, as in the first problem. Substituting it into Eq. (10) gives different power law prefactors, $l^{-2}$ in the left-hand side and $l^{-1}$ in the right-hand side. This defect can be readily removed by replacing the constant $a$ by a slowly varying function of $l$. An asymptotic analysis yields

$$p(l) \sim \exp\left(-\frac{\ln(1/l) + \ln\ln(1/l) + \ldots}{l}\right). \quad (14)$$

Our simulational results well agree with the asymptotic prediction of Eq. (14).

The singularities seem to be a generic feature of the random cutting model and the likes. It may be difficult to prove generally, but we believe that the mere existence of singularities and their location are robust to small changes in the model rules. In particular, if one considers not necessarily uniform cutting probability, $\rho(x) \neq \text{const}$, one finds that the present approach still applies and only quantitative changes appear, e.g., integral equations will contain the cutting probability $\rho(x)$. To verify the robustness of generic properties we consider a $d$-dimensional random cutting model. For example, in two dimensions the model is defined as follows. We start with a unit square, pick an arbitrary internal point and cut the square along the two lines going through this point in $x$- and $y$-directions. We then continue the cutting process with only one of four rectangles, say with the north-east one. We wish to determine the probability density, $p(A)$, for the rectangle of the largest area $A$ which arises in the cutting process.

Note first that for the uniform cutting probability, $\rho(x,y) = 1$, the area distribution function is not uniform even in the one-cut process. Instead, it is given by

$$p_0(A) = \int_0^1 \int_0^1 dx\,dy\, \delta(xy - A) = -\ln(A). \qquad (15)$$

The probability density satisfies an integral equation

$$p(A) = 3p_0(A) + \int_0^1 \int_0^1 dx\,dy\, \theta(xy - A) \frac{1}{xy}\, p\left(\frac{A}{xy}\right), \qquad (16)$$

where $\theta$ is the step function and the factor 3 in the first term accounts for three rectangles which are kept unchanged in the following cutting process. Eq. (16) is derived in the same way as Eq. (7) in the corresponding $1D$ version and valid for $A \geq \frac{1}{2}$, again like in the $1D$ situation. Changing variables from $(x,y)$ to $(u = xy, v = x/y)$, performing the $v$-integration, twice differentiating the remaining integrals, and solving the resulting differential equation, we find an exact solution,

$$p(A) = \frac{3}{2}\left(\frac{1}{A} - A\right), \quad \text{for} \quad A \geq \frac{1}{2}. \qquad (17)$$

This result perfectly agrees with simulations (Fig. 3).

Simulations in two dimensions also clearly indicate singularities at $A = \frac{1}{2}$ and $\frac{1}{4}$. However, we still expect the singularities at all inverse integers, $A = \frac{1}{k}$. The qualitative origin of singularities is the same as previously. If in the cutting process a sufficiently large rectangle, namely a rectangle of area $A \geq \frac{1}{2}$ is formed, it is always the largest. For the largest rectangle of a smaller area $A < \frac{1}{2}$ we should additionally guarantee that other rectangles are smaller than $A$ and hence the probability distribution function $p(A)$ is described by different expressions for $A \geq \frac{1}{2}$ and $A < \frac{1}{2}$. When we pass through other critical points $A = \frac{1}{k}$ new possibilities arise and hence the probability distribution function changes again. Geometry comes into play and changes the strength of singularities, but neither their existence nor location. Indeed, in $2D$ one can form two rectangles of the area $A = \frac{1}{2}$ or four rectangles of the area $A = \frac{1}{4}$ just in one cut. However, one cannot form three rectangles of the area $A = \frac{1}{3}$ after one cut – one needs an additional cut. Note that a singularity which arises at the later stage of the cutting process is weaker – mathematically it appears by "integrating" the previous singularity. In one dimension the strengths of singularities were consecutive, see Eqs. (6) and (13), since they appear on the consecutive stages of the cutting process. In $2D$ both singularities at $A = \frac{1}{2}$ and $\frac{1}{4}$ arise after the first stage and thence identical (simulations indicate that $p_{\text{sing}}(A) \sim |A - \frac{1}{2}|$ and $p_{\text{sing}}(A) \sim |A - \frac{1}{4}|$). The singularity at $A = \frac{1}{3}$ appears later and should be weaker. Unfortunately, we could not find its strength neither theoretically nor numerically.

We believe that the largest volume distribution shows the same properties in all dimensions. We have verified this in three dimensions. The singularities become weaker as the dimension grows but we have certainly detected first few of them.

In summary, we have demonstrated surprisingly peculiar properties of the longest-intervals probability distribution. We have found an infinite set of singularities at the inverse integer values of the the longest interval $l = \frac{1}{k}$. Singularities become weaker and weaker as $k$ increases. Additionally, we have found an essential singularity at $l = 0$. We have confirmed the existence of these singularities for several models. For the random cutting model, these results are valid in arbitrary dimension. We expect these singularities to appear in other apparently unrelated problems.

We are thankful to S. Redner for interesting discussions. We gratefully acknowledge support from the Swiss National Foundation (to L.F.), from ARO through grant DAAH04-93-G-0021 and from NSF through grant DMR-9219845 (to I.I. and P.L.K.).